\documentclass[%
twocolumn%
,secnumarabic%
,amssymb, amsmath,nobibnotes, aps, showpacs]{revtex4}
\usepackage{bm}%
\newcommand{\rmd}{\mathrm{d}}
\newcommand{\rmi}{\mathrm{i}}
\newcommand{\rme}{\mathrm{e}}
\newcommand{\Tr}{\operatorname{Tr}}
\newcommand{\EE}{\operatorname{\mathbb{E}}}
\newcommand{\Qbb}{\mathbb{Q}}
\newcommand{\Pbb}{\mathbb{P}}
\begin{document}
\title{Stochastic Schr\"odinger equations with coloured noise}
\author{A. Barchielli}
\email{Alberto.Barchiellli@polimi.it} \affiliation{Politecnico di Milano, Dipartimento di
Matematica \\ and Istituto Nazionale di Fisica Nucleare, sezione di Milano}
\author{C. Pellegrini}
\email{clement.pellegrini@math.univ-toulouse.fr}
\affiliation{Laboratoire de Statistique et Probabilit\'es, Universit\'e Paul Sabatier,\\ 118, Route de Narbonne, 31062 Toulouse Cedex 4,
France}

\author{F. Petruccione}
\email{petruccione@ukzn.ac.za}
\affiliation{Quantum Research Group, School of Physics and National Institute for Theoretical Physics,\\
University of KwaZulu-Natal, Private Bag X54001,\\Durban 4000, South Africa}

\begin{abstract}
A natural non-Markovian extension of the theory of white noise quantum trajectories is presented. In order to introduce memory effects in the formalism an Ornstein-Uhlenbeck coloured noise is considered as the output driving process. Under certain conditions a random Hamiltonian
evolution is recovered.  Moreover,  non-Markovian stochastic Schr\"odinger equations which unravel non-Markovian master equations are derived.
\end{abstract}
\maketitle

Recently, stochastic wave function methods for the description of open quantum systems have
received considerable attention \cite{CarmichaelBooks,GardinerZoller,Book}. These approaches have been mainly motivated by the continuous
measurement description of detection schemes in quantum optics, namely direct photo detection,
homodyne and heterodyne detection. In general, selective, indirect and continuous quantum
measurements allow for a description of the open quantum system in terms of the stochastic
evolution of its own wave function. Typically, the random trajectories of the state vector
involve quantum jumps or diffusion processes and are described by stochastic differential
equations for the wave function $|\psi_t\rangle$.

The relationship to the more traditional approach to open quantum systems in terms of Master equations for the density
matrix $\rho_t$ is easily established by realizing that the latter can be expressed as $\rho_t
= \EE[|\psi_t\rangle \langle \psi_t|]$, where $\EE$ denotes the ensemble average over
realizations of the stochastic process $|\psi_t\rangle$. Thus, the Master equation evolution
can be reproduced by generating a large number of trajectories of the state vector. This
\textit{unravelling} procedure, called Monte Carlo wave function method, has gained
considerable importance for the numerical simulation of complex open systems \cite{Book}.

The procedure described above has been a major breakthrough in the description of the
Markovian dynamics of open quantum systems, where it has found many applications. At present
an active line of research is to find generalizations of the stochastic Schr\"odinger equation
in the non-Markovian regime, which have a sound physical interpretation.

Essentially, two strategies have been investigated to accomplish this goal. A first strategy is to
start from non-Markovian master equations and to try to construct pure state unravellings.  It
is well-know that non-Markovian master equations can be constructed with the help of
projection operator techniques. For example, time-convolutionless (TCL) projection techniques
yield master equations that are local in time and allow for suitable unravellings
(\cite{Book,Piilo} for details). In the same spirit, non-Markovian generalizations of the
Lindblad theory \cite{BreuerPRA2007} have been proposed and shown to have corresponding
non-Markovian unravellings \cite{Merve,OtherGuy}. The measurement interpretation of these
unravellings is still an open problem. However, special cases of unravellings with a clear
measurement interpretation have been shown to exist for discrete time models of repeated
interaction \cite{ClementFrancesco}.

The second strategy is first to generalize directly the stochastic Schr\"odinger equation and
secondly to show if it provides an unravelling of a suitable Master equation and if it has a
measurement interpretation. In \cite{Diosi,Wiseman,Strunz,Salgado,Piilo,ClementFrancesco}
starting from a measurement scheme a rigourous non-Markovian random evolution has been
derived. The question of the possibility to express this random evolution in terms of pure
states in the spirit of unravelling is still highly debated \cite{Wiseman,Diosi}. For the
usual scheme of indirect quantum measurement it has been shown that in general such an interpretation is
not accessible \cite{Wiseman,ClementFrancesco}. Other investigations promise to get unravelling 
\cite{Diosi,others,ClementFrancesco}.

In classical statistical physics the most straightforward way to introduce memory effects is
to consider coloured noise \cite{Coloured_noise}. This approach has already been successfully
transferred to the theory of isolated quantum systems, by introducing coloured noise in the
Hamiltonian of  the evolution \cite{Coloured_noise,Kubo,Wodkieski,Eberly}. In this paper we show how
to introduce memory effects in the stochastic Schr\"odinger equation with the help of the
addition of coloured noise. Specifically, we will illustrate the approach by using the 
Ornstein-Uhlenbeck process.

The alternative strategy, proposed here, is focused on an extension of the canonical Markovian
approach. To be precise, generic Markovian stochastic Schr\"odinger equations are justified by
the use of a linear equation for a non normalized state $\psi_t$ of the following form
\begin{equation}\label{lin1}
\rmd \psi_t=A\psi_t\rmd t+B\psi_t\rmd X_t,
\end{equation}
where the formal derivative of $(\dot X_t)$ is a white noise, that is, $\EE[\dot X_t\dot
X_s]=\delta(t-s)$ and $\EE[X_t]=0$ ($\EE$ denotes the expectation under a reference
probability). In general, the process $(X_t)$ can be a Brownian motion, a counting process or
a mixing of both noises. Physically, the noise terms represent the output signal of the
measurement and the probability of an event is traditionally  described by
$\Vert\psi_t\Vert^2$. Usually, asking $\Vert\psi_t\Vert^2$ to represent a probability density
(giving the physical probability) implies $\EE[\Vert\psi_t\Vert^2]=1$ and this last equality
imposes the form of the operators $A$ and $B$. Finally,  a new stochastic differential
equation for normalized vector states $(\hat{\psi}_t)$ is derived. In the equation for the
normalized states a new white noise appears, obtained by a suitable Girsanov transformation of
the reference Brownian motion. A natural introduction of non-Markovian memory effects in the
framework sketched above, can be obtained by replacing the white noise $(X_t)$ by a coloured
Ornstein-Uhlenbeck noise.

To keep things simple here we consider a one-dimensional Ornstein-Uhlenbeck process defined by
$X_t = \int_0^t \rme^{-\gamma (t-s)} \rmd W_s$, where $(W_t)$ is a one dimensional Brownian
motion, under a reference probability $\mathbb{Q}$.  The Ornstein-Uhlenbeck process $(X_t)$
satisfies the stochastic differential equation
\begin{equation}\label{eqor}
\rmd X_t= -\gamma X_t \rmd t +
\rmd W_t,
\end{equation}
where $\gamma>0$. The high non-Markovianity of the process is emphasized by the fact that the correlation
function 
is not a pure delta function, but we have $\EE_\mathbb{Q}[\dot X_t\dot X_s]=\delta(t-s) -
\frac \gamma 2 \left(\rme^{-\gamma\vert t-s\vert}+\rme^{-\gamma(t+s)}\right)$ and the
Markovian regime is recovered in the limit $\gamma \downarrow 0$.

In the following, first we will  show how this approach allows to reconstruct dynamical models
with random Hamiltonians. Next, we will demonstrate how to construct a class of non-Markovian
stochastic Schr\"odinger equations, that has an interpretation in terms of quantum measurement
and unravelling.

\paragraph*{Random Hamiltonians.}
Actually, the random Hamiltonian models arise naturally by translating the action of the
Ornstein-Uhlenbeck process \eqref{eqor} in the definition of the linear stochastic
Schr\"odinger equation  \eqref{lin1}. With the output driving process $(X_t)$ satisfying
\eqref{eqor},   Eq.\ \eqref{lin1} can be rewritten in the following form
\begin{equation}\label{lin2}
\rmd \psi_t=(A-\gamma X_tB)\psi_t\rmd t+ B\psi_t\rmd W_t,
\end{equation}
where the initial condition is a wave function $\psi_0$, such that $\Vert\psi_0\Vert^2=1$.

According to the laws of quantum measurement the probability of the output is described by the
density $\Vert\psi_t\Vert^2$. This yields a measurement interpretation of Eq.\ \eqref{lin2}.
The quantity $\Vert\psi_t\Vert^2$ defines a probability density, if
$\EE_\Qbb[\Vert\psi_t\Vert^2]=1$, is satisfied for all times $t$. The consistency of these
densities needs also the process $(\Vert\psi_t\Vert^2)$ to be a martingale (we refer to
\cite{BarGreg09} for complete details). By It\^o calculus rules, the stochastic differential of
$\Vert\psi_t\Vert^2$ turns out to be
\begin{multline}\label{diffnormsq}
\rmd\langle\psi_t|\psi_t\rangle=\langle
\rmd\psi_t|\psi_t\rangle+\langle\psi_t|\rmd\psi_t\rangle+
\langle \rmd\psi_t|\rmd\psi_t\rangle \\
{}=\langle\psi_t|\left[A^*+A-\gamma X_t\left(B^*+B\right)+ B^*
B\right]\psi_t\rangle\rmd t \\
{}+\langle\psi_t|\big( B^*+ B\big)\psi_t\rangle\rmd W_t.
\end{multline}
In order to ensure that $\EE_\mathbb{Q}[\Vert\psi_t\Vert^2]=1$, for all times $t$, we must impose
that the term in $\rmd t$ must be equal to zero (and this guarantees also the martingale
property). Indeed, in this way, we get
$\EE_\mathbb{Q}[\Vert\psi_t\Vert^2]=\EE_\mathbb{Q}[\Vert\psi_0\Vert^2]=1$, since we have
$\EE_\Qbb[\rmd W_t]=0$.  It follows that we must have
\begin{equation}\label{cond1}A^*+A-\gamma X_t\left(B^*+B\right)+B^* B=0,\qquad \forall t.
\end{equation}
This imposes that there are two self-adjoint operators $K$ and $H$ such that $B=-\rmi K$ and
$A=-\rmi H-\frac{1}{2}K^2$. Then, Eq.\ \eqref{lin2} becomes
\begin{equation}\label{liin3}
\rmd\psi_t=\left[-\rmi\left(H-\gamma X_tK\right)-\frac{1}{2}\,K^2\right]\psi_t \rmd t
-\rmi K\psi_t\rmd W_t\,.
\end{equation}

Let $\textrm{T}_\leftarrow$ denotes the time ordering exponential, the formal solution of Eq.\ \eqref{liin3} is given by
\[
\psi_t=\textrm{T}_\leftarrow\left\{-\rmi \int_0^t\left(H-\gamma X_sK\right) \rmd s
-\rmi \int_0^t K \rmd W_s \right\}
\psi_0.
\] In other terms, the evolution of the quantum system is completely determined by the
time-dependent, random Hamiltonian $\hat{H}_t=H+\left( \dot W_t-\gamma X_t \right) K$ (a
formal expression, due to the presence of $\dot W_t$). Note that Eq.\ \eqref{diffnormsq} gives
$\rmd \|\psi_t\|^2=0$, in agreement with a purely Hamiltonian evolution. This shows, that the
usual measurement interpretation of \eqref{lin1} \textit{coloured} with an Ornstein-Uhlenbeck
process gives raise to a random Hamiltonian evolution. The property $\|\psi_t\|^2=1$ implies
that we have  not extracted information according to our measurement interpretation.  We
recover the framework of the evolution of an isolated closed system incorporating a random
environment characterized in terms of Ornstein-Uhlenbeck noise.

In order to investigate the property of Eq.\ \eqref{liin3} in terms of unravelling, it is
interesting to consider the evolution of the corresponding density matrices. To this end, we
consider the pure state process $(\rho_t)$ defined by $ \rho_t=|\psi_t\rangle \langle
\psi_t|$. We define also the mean state $\eta_t=\EE_\Qbb[\rho_t]$. Using It\^o rules, the
process $(\rho_t)$ satisfies the stochastic differential equation (SDE)
\begin{multline}\label{eqrho1}
\rmd \rho_t= -\rmi [H-\gamma X_tK,\rho_t]\rmd t -\rmi  [K,\rho_t]\rmd W_t
\\ {}
-\frac {1} 2 \,[K,[K,\rho_t]]\rmd t,
\end{multline}
which is, of course, equivalent to \eqref{liin3}. Let us stress that the presence of the
Ornstein-Uhlenbeck process implies that the solution $(\rho_t)$ of Eq.\ \eqref{eqrho1} is not
a Markov process. For the mean state $(\eta_t)$ we have then
\begin{equation}\label{meqmem}
\frac{\rmd\ }{\rmd t}\, \eta_t= -\rmi [H,\eta_t] -\frac {1} 2 \,[K,[K,\eta_t]]
+\rmi \gamma\big[K,\EE_\Qbb[X_t\rho_t]\big].
\end{equation}
The above equation can be naturally considered as a  Master equation, but its particularity is
that it is not a closed equation for the mean state $\eta_t$. Actually, we have derived a
model with memory for the mean state. Indeed, the term $\rmi\gamma
\big[K,\EE_\Qbb[X_t\rho_t]\big]$ introduces  non-Markovian memory effects in the dynamics.
Moreover, Eq.\ \eqref{liin3} is an unravelling of the Master equation \eqref{meqmem}.



\paragraph*{Random coefficient}
The random Hamiltonian case we discussed above arose only because the coefficients $A$ and $B$
in the Eq.\ \eqref{lin1} were assumed to be not random. In the following we will address the
situation, in which  $A$ and $B$ are assumed to be functions of the Ornstein-Uhlenbeck process
$(X_t)$, thus becoming random operators themselves. In other words, we consider now a SDE of
the more general form
\begin{equation}\label{lin3}
\rmd\psi_t=A(X_t)\psi_t\rmd t+B(X_t)\psi_t\rmd X_t.
\end{equation}
Inserting the definition \eqref{eqor} of the Ornstein-Uhlenbeck process into Eq.\ \eqref{lin3}
yields
\begin{equation}\label{ABX}
\rmd\psi_t=\big(A(X_t)-\gamma X_tB(X_t)\big)\psi_t\rmd t+ B(X_t)\psi_t\rmd W_t.
\end{equation}
Again, asking the norm $\Vert \psi_t \Vert^{2}$ to be a probability density the condition
\eqref{cond1} is translated into the following one
\begin{multline}
\label{cond2}
A^*(X_t)+A(X_t)-\gamma X_t\big(B^*(X_t)+B(X_t)\big) \\ {}
+B^*(X_t) B(X_t)=0.
\end{multline}
In contrast to the previous case with deterministic coefficients $A$ and $B$, we can now
establish a relationship between the operators $A(X_t)$ and $B(X_t)$. More precisely, the
condition \eqref{cond2} implies that there exists a self-adjoint operator $H(X_t)$ such that
\[
A(X_t)-\gamma X_tB(X_t)=-\rmi H(X_t)-\frac{1}{2}\,B^*(X_t)
B(X_t).
\]
Then, Eq.\ \eqref{ABX} becomes
\begin{multline} \label{lin33}
\rmd \psi_t=\left(-\rmi H(X_t)-\frac{1}{2}\,B^*(X_t) B(X_t)\right)\psi_t\rmd t \\ {}+
B(X_t)\psi_t\rmd W_t.
\end{multline}
In the general case $\rmd \Vert \psi_t \Vert \neq 0$ and the above evolution cannot be
interpreted in terms of a random Hamiltonian. This is the signature that the above evolution
describes the extraction of information from the system in terms of indirect quantum
measurement. Indeed, the measurement interpretation relies upon the property that the norm
$\Vert \psi \Vert^{2}$ defines a probability density. According to this property the relevant
physical probability is defined by
\[
\mathbb{P}^T(\rmd \omega)=\Vert\psi_T(\omega)\Vert^2\mathbb{Q}(\rmd \omega),
\]
where $\mathbb{Q}$ is the underlying reference probability as introduced in the definition of
the Ornstein-Uhlenbeck process. In other words, the probability $\mathbb{P}^T$ is the physical
probability law of the events which could occur in the time interval $[0,T]$. When the time
$T$ varies, the probabilities $\mathbb{P}^T$ are ``consistent'', because $\Vert\psi_T\Vert^2$
is a ``martingale'' \cite{BarGreg09}. Consistency means that if we take $0<S<T$ and an event
determined by conditions only in the time interval $(0,S)$, then
$\mathbb{P}^T(F)=\mathbb{P}^S(F)$. Another important property of these probabilities is that
they can be expressed in terms of quantum expectations on the premeasurement state of positive
operator valued measures, as stated by the axioms of quantum mechanics \cite{BarGreg09}.

In order to describe the stochastic dynamics of the system undergoing continuous indirect
measurement we derive the corresponding stochastic Schr\"odinger equation for the normalized
state vector $\widehat{\psi}_t=\psi_t/\Vert\psi_t\Vert$. Usually, such a process is called a
\textit{quantum trajectory} and satisfies a non-linear stochastic differential equation. To
derive this equation we need to describe the driving random process in terms of  the new
physical probability $\mathbb{P}^T$. The famous Girsanov theorem gives that, under
$\mathbb{P}^T$, the process
\[
\widehat{W}_t=W_t-\int_0^tm(s)ds,
\]
with
\[
m(t)=\big\langle\widehat{\psi}_t\big|\big( B^*(X_t)+ B(X_t)\big)
\widehat{\psi}_t\big\rangle,
\]
is a new Brownian motion. This allows to express the output of the measurement $X_t$ (which
was an Ornstein-Uhlenbeck coloured noise under the reference probability $\Qbb$) in the form
\[
X_t=\int_0^t \rme^{-\gamma (t-s)}m(s)\rmd s +\int_0^t \rme^{-\gamma (t-s)}\rmd \widehat W_s.
\]

Moreover, by using It\^o calculus we can write the non-linear stochastic differential equation
for the normalized state vector $\widehat{\psi}$ as
\begin{multline}
\label{nmsse}
\rmd\widehat{\psi}_t=\left[B(X_t)-\frac{m(t)}{2}\right]\widehat{\psi}_t \rmd\widehat{W}_t -
\biggl[\rmi H(X_t)+\frac{m(t)^2}{8}\\ {}- \frac{m(t)}{2}\, B(X_t)+\frac{1}{2}\,B^*(X_t)B(X_t)
\biggr]\widehat{\psi}_t\rmd t.
\end{multline}
Let us stress that the presence of the coloured noise $X_{t}$ in the above equation makes the process
$\widehat{\psi}$ non-Markovian. Thus, Eq.\ \eqref{nmsse} is an example of a non-Markovian stochastic
Schr\"odinger equation that admits a clear interpretation in terms of indirect quantum measurement.
Eq.\ \eqref{nmsse} has the usual structure of a stochastic Schr\"odinger equation.
Its particularity is the fact that the operators $H$ and $B$ are now random functions of
the Ornstein-Uhlenbeck process $X_{t}$.

We shall investigate the evolution of the corresponding density matrix. To this end we define
$$
\widetilde{\rho}_{t} = \vert \widehat{\psi}_{t}\rangle \langle  \widehat{\psi}_{t} \vert.
$$
The non-linear stochastic master equation for the density matrix $\widetilde{\rho}_{t}$ takes the form
\begin{multline}
\rmd\widetilde\rho_t=\mathcal{L}(X_t)[\widetilde\rho_t]\rmd t+\Big[ B(X_t) \widetilde\rho_t+
\widetilde\rho_t B^*(X_t)\\
{}-\Tr\left\{\left[ B(X_t)+
B^*(X_t)\right]\widetilde\rho_t\right\}\widetilde\rho_t\Big]\rmd\widehat{W}_t\,,
\end{multline}
where
\begin{multline*}
\mathcal{L}(X_t)[\rho]=-\rmi[H(X_t),\rho]\\ {}-\frac{1}{2}\{B^*(X_t)B(X_t), \rho\}+B(X_t)\rho
B^*(X_t),
\end{multline*}
defines a random Lindblad generator. Note, that again the originality relies on the fact that
the operators $H$ and $B$ are random (see \cite{Budini,Salgado} for similar considerations). 

The mean state $\eta_t=\EE_{\Pbb^T}[\widetilde\rho_t]=\EE_{\Qbb}[|\psi_t\rangle \langle
\psi_t|]$ satisfies the following non-closed equation
\begin{equation}
\label{nceq}
\frac{\rmd\ }{\rmd t}\, \eta_t=\EE_{\Pbb^T}\big[\mathcal{L}(X_t)[\widetilde\rho_t]\big].
\end{equation}
This equation describes a non-Markovian deterministic Master equation. Of course, by
construction the process $\widehat{\psi}_{t}$ yields a non-Markovian unravelling of the Master
equation \eqref{nceq}.

A special case is featured by considering $B$ to be constant. In this case, we recover a model
for the open system with a random Hamiltonian and a usual dissipative part in Lindblad form
with a non random operator $B$.
\smallskip

In conclusion, a special class of non-Markovian stochastic Schr\"odinger equations and
stochastic master equations has been derived. The framework of the traditional
\textit{Markovian white noise} theory has been extended by introducing the use of coloured
Ornstein Uhlenbeck noise. This natural extension allows to describe random dynamics for open quantum system with memory effects. Two important categories of non-Markovian models has been derived. On the one hand, in a natural way, we have recovered the special case of random Hamiltonians. On the other hand, a generalization of Lindblad type equations has been obtained. The memory effects, influenced by the Ornstein Uhlenbeck noise, are encoded in the definition of the random operators which define either the Hamiltonian or the Lindblad operator. Our approach, based on the description of stochastic Schr\"odinger equations in terms of pure states, give rises to non-Markovian unravellings of these models.

\begin{acknowledgments}
This work is based upon research supported by the South African Research Chairs Initiative of the Department of Science and Technology and National Research Foundation.
\end{acknowledgments}

\end{document}